# Impact of high-pressure torsion on hydrogen production from photodegradation of polypropylene plastic wastes

Thanh Tam Nguyen[a,b] and Kaveh Edalati[a,b],*

[a] WPI, International Institute for Carbon Neutral Energy Research (WPI-I2CNER), Kyushu University, Fukuoka 819-0395, Japan
[b] Mitsui Chemicals, Inc. -.Carbon Neutral Research Center (MCI-CNRC), Kyushu University, Fukuoka 819-0395, Japan

Plastic waste entering the environment through landfilling or improper disposal poses substantial risks to ecosystems and human health. Photoreforming is emerging as a clean photocatalytic technology that degrades plastic waste to organic compounds while simultaneously producing hydrogen fuel. This study introduces high-pressure torsion (HPT), a severe plastic deformation (SPD) method, as an innovative technique to enhance the photoreforming of polypropylene (PP) plastic mixed with a brookite $TiO_2$ photocatalyst. Hydrogen production systematically increases with the number of HPT turns, accompanied by the formation of valuable small organic molecules. The enhancement in photocatalytic activity is attributed to strain-induced defect formation in both catalysts and plastics, as well as the creation of catalyst/plastic interphases that enhance charge carrier transport between inorganic and organic phases. These findings reveal a new functional application for SPD in energy conversion and sustainability.

*Corresponding author (E-mail: kaveh.edalati@kyudai.jp; Tel: +80-92-802-6744)

## 1. Introduction

The global consumption of plastics continuously rises due to extensive human activities, becoming inevitable as plastics are integral to nearly every economic sector [1]. Annually, millions of tons of thermoplastic polymers such as polypropylene (PP), polyethylene (PE), polyvinyl chloride (PVC) and polyethylene terephthalate (PET) are manufactured [2-5]. Despite their advantageous mechanical properties, thermoplastics have notably short useful lifespans, and their disposal poses significant environmental challenges due to their low degradation rates [6,7]. Among thermoplastics, PP, which is favored for its low production cost, excellent thermal and chemical resistance, high tensile strength, and relatively low density [8,9], is utilized in producing various items including bottles, toys, automotive parts, medical devices, and fabrics [10-12]. However, PP as a non-biodegradable thermoplastic polymer exhibits a slow degradation rate, leading to its accumulation in landfills worldwide without being recycled [9].

Several methods have been investigated to mitigate plastic waste, including thermal degradation [13], incineration [14], landfilling [15] and ozonation [16]. However, these approaches are often energy-intensive and costly. Recent investigations have focused on alternative techniques such as biodegradation [17] and photocatalysis [18]. Biodegradation involves microbes producing enzymes that break down macromolecules into small fragments, potentially leading to the complete mineralization of plastic wastes [19]. Photocatalysis is a promising and efficient process that uses only sunlight to degrade a wide range of organic pollutants into $CO_2$ and $H_2O$ [20]. A modern advancement in photocatalysis is photoreforming, schematically shown in Fig. 1a, which involves the simultaneous reduction of water and the oxidation of organic materials [21-23]. The photoreforming process can particularly provide a clean pathway for upcycling of plastic wastes [24-26]. This process uses plastic waste as a feedstock for clean hydrogen production and small organic molecule generation under mild processing conditions [27-32].

Titanium dioxide ($TiO_2$) is recognized for its exceptional semiconducting, optical, and catalytic properties, making it an effective and stable photocatalyst [33-35]. $TiO_2$ has three natural crystalline polymorphs including anatase (tetragonal), brookite (orthorhombic), and rutile (tetragonal) [36]. A previous study demonstrated that brookite exhibits the highest activity for the photoreforming of PET plastic due to easy hydroxyl radical formation, moderate depth of electron-trap states, slow decay rate, and efficient charge transfer kinetics [37]. Despite this report, there have been no attempts to use brookite in the photoreforming of hard-to-degrade plastics such as PP.

In this study, brookite $TiO_2$ is used for simultaneous PP plastic degradation and hydrogen production. Due to the low efficiency of photoreforming for PP, severe plastic deformation (SPD) via the high-pressure torsion (HPT) method [38,39] is used to mix PP and brookite and successfully enhance the activity. HPT is a relatively new dopant-free approach in the field of catalysis that is based on mechanical processing [39]. In contrast to HPT, most other dopant-free approaches applied to $TiO_2$ photocatalysts, such as the introduction of titanium interstitials [40], enrichment of oxygen [41], reduction of oxygen to create black titania [42], and polymorphic transformations like two-dimensional phase production [43], are based on chemical processing.

## 2. Experimental Procedure

Homopolymer PP (h-PP) powder with spherical particles and sizes of less than 63 μm was supplied by Mitsui Chemicals, Inc., Japan, while brookite with a purity of 99.99% was obtained from Kojundo Chemical Company, Japan. Examination of brookite powder by scanning electron microscopy (SEM, as shown in Fig. 1b) indicated an average particle size of 2.6 μm, and the presence of numerous crystals within these microparticles was confirmed by transmission electron microscopy (TEM, as shown in Fig. 1c).

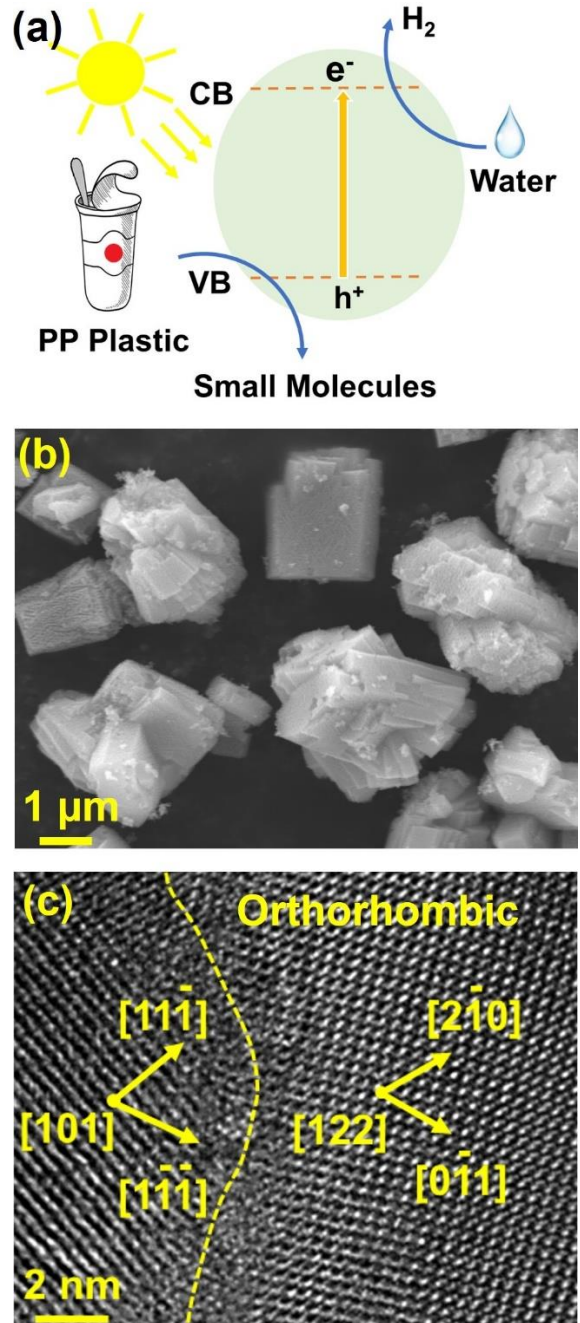

Figure 1. (a) Schematics of photoreforming process for PP plastic, (b) morphology of brookite examined by SEM, and (c) examination of crystals in brookite by TEM.

A mixture of 50 wt% brookite and 50 wt% PP plastic was prepared in acetone using a mortar and pestle for 30 min. This blended powder was then pressed into a disc with a 10 mm diameter under a pressure of 380 MPa. The disc underwent mechanical processing via the HPT method at a pressure of $P = 6$ GPa at ambient temperature, with a rotation speed of 1 rpm for $N =$ 1 (moderate external strain) and 3 turns (strong external strain). After HPT processing, the brookite and PP mixture, which was in the form of a disc, was manually crushed using a mortar and pestle.

The microstructure of the HPT-processed samples was examined by SEM coupled with energy-dispersive X-ray spectroscopy (EDS) under an acceleration voltage of 5 keV. The crystalline structures were analyzed by X-ray diffraction (XRD) with Cu Kα irradiation. Attenuated total reflectance Fourier transform infrared spectroscopy (FTIR) within the 600-4000 cm$^{-1}$ wavenumber range was used to investigate the molecular composition and structure of the PP plastic. Steady-state photoluminescence (PL) measurements with a 325 nm laser source were performed to assess the radiative electron-hole recombination.

For photoreforming experiments, 100 mg of powder mixtures (containing 50 mg of brookite and 50 mg of PP) before and after HPT processing were added to 3 mL of 10 M NaOH solution. Subsequently, 250 µL of 0.01 M $Pt(NH_3)_4(NO_3)_2$ was added, providing platinum as a co-catalyst at 1 wt% of the brookite catalyst. This mixture was sonicated for 5 min, air-evacuated with argon for 30 min, and then irradiated under continuous stirring with the full arc of a 300 W xenon lamp at an intensity of 18 kW/m², maintaining a constant temperature of 25 °C. The amount of produced hydrogen was measured using gas chromatography (GC) equipped with a thermal conductivity detector.

The degradation products of PP plastic were identified using a gas chromatography - mass spectrometry (GC-MS QP2010, Shimadzu, Japan) device equipped with a fused silica capillary column coated with CP-SIL 8 CB. Before GC-MS analysis, liquid samples after the photocatalytic tests were pre-treated using the solid phase extraction (SPE) technique with crosslinked poly(styrene divinylbenzene) (ENV+) columns. The activation process for SPE columns involved using 1 mL of methanol followed by 1 mL of deionized water. Then, 20 mL of the post-photocatalysis sample was slowly loaded through the SPE column. Next, 10 mL of methanol was used to elute the extracted substances. The collected samples were evaporated under argon gas, reducing the volume from 10 to 2 mL. The GC oven was initially held at 80 °C for 5 min, then heated to 270 °C at a rate of 50 °C/min, and maintained at 270 °C for 3 min. The transfer line and MS ion source temperatures were set to 280 °C and 230 °C, respectively. An aliquot of 1 µL of the extracted sample was injected into the GC injector, maintained at 250 °C, and MS-EI was employed to screen the degradation products over a mass-to-charge (*m/z*) range of 30-550, with helium serving as the carrier gas for the analysis.

## 3. Results

Fig. 2 shows the SEM and SEM-EDS images of brookite and PP plastic mixed using HPT for $N = 3$ turns. As shown in Fig. 2a, using SEM back-scatter electron mode, and in Fig. 2b-d, using EDS elemental mapping, PP and brookite are significantly mixed. Neither PP nor brookite retain their initial shapes, indicating the occurrence of deformation during the process. In addition to mechanical mixing, which results in the formation of brookite-PP interphases, a large number of cracks are observed in PP, as shown in higher magnification in Fig. 2e. Moreover, a large amount of nano-sized brookite is detected on PP in Fig. 2e. This indicates that the brookite micropowders are partly fragmented during HPT processing. This observation is inconsistent with earlier publications on the application of HPT to different ceramics [38,39], including brookite [44], which reported powder consolidation rather than fragmentation. This difference is due to the presence of PP, which forms a composite with brookite. It is known that components in composites can be fragmented by HPT processing due to constraints in their co-deformation [45]. The formation of nano-sized brookite from the initial brookite micropowder along with the large contact area between brookite and PP, are beneficial for photoreforming.

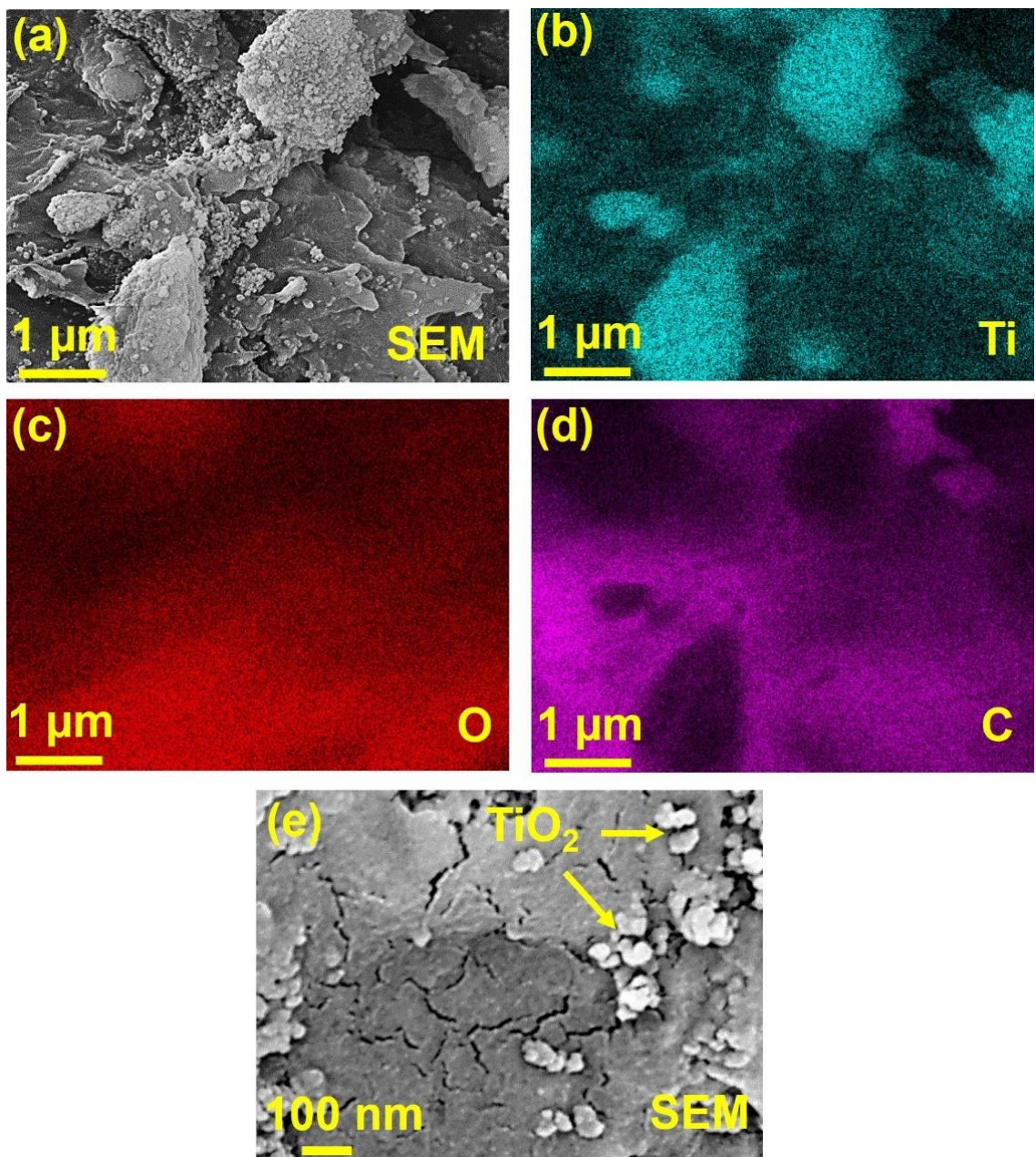

Figure 2. (a) SEM image in secondary electron mode and (b-d) corresponding EDS mappings, and (e) SEM image in backscatter electron mode for brookite-PP powder mixture processed by HPT for $N$ = 3 turns.

Fig. 3a shows the XRD profiles of the starting brookite without HPT processing and the brookite-PP mixture processed by HPT with $N$ = 1 and 3 turns. Brookite exhibits the orthorhombic structure with the Pbca space group in all samples. The lattice parameters are $a$ = 9.1740 nm, $b$ = $c$ = 5.4490 nm, and $\alpha = \beta = \gamma = 90°$, which are in good agreement with the JCPDF 01-076-1936 card for brookite. No phase transformation is detected after HPT processing, but peak broadening occurs, indicating lattice strain caused by the formation of linear and planar defects by HPT [38,39,44,45]. Such defects can act as active sites for enhancing photocatalytic activity.

The FTIR spectrum of the initial PP plastic without HPT processing and the brookite-PP mixture processed by HPT with $N = 1$ and 3 turns are shown in Fig. 3b. The FTIR peaks of PP plastic include isotactic at around 800-1300 cm$^{-1}$, -CH$_3$ bend at 1377 cm$^{-1}$, -CH$_2$ bend at 1458 cm$^{-1}$, and C-H stretch at 2700-3000 cm$^{-1}$, which are consistent with previous reports [46-48]. After mixing with brookite using HPT, the peak intensities of PP decrease, and the TiO$_2$ bands [49] appear in the FTIR spectra. The FTIR spectra suggest that HPT does not lead to detectable molecular structure changes in PP.

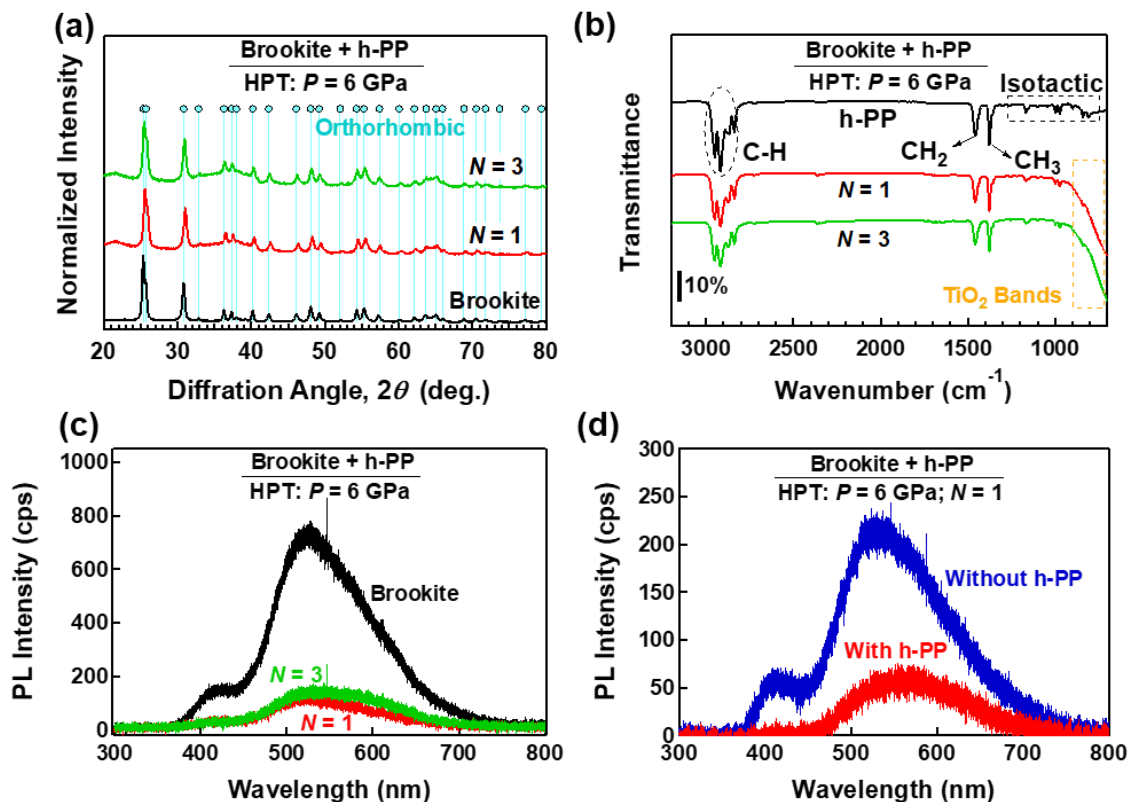

Figure 3. (a) XRD spectra of initial brookite without HPT and brookite-PP mixture after $N = 1$ and 3 HPT turns, (b) FTIR spectra of initial PP plastic without HPT and brookite-PP mixture after $N = 1$ and 3 HPT turns, (c) PL spectra of initial brookite without HPT and brookite-PP mixture after $N = 1$ and 3 HPT turns, and (d) PL spectra of brookite and brookite-PP mixture after HPT processing for $N = 1$ turns.

The radiative recombination of photo-induced electrons and holes was examined by steady-state PL (Fig. 3c and 3d). In the brookite PL spectrum (Fig. 3c), a PL peak at around 410 nm corresponds to band-to-band electron-hole recombination, and another peak at a higher wavelength around 520 nm corresponds to recombination on defects [37]. The PL intensities in the mixture samples of PP and brookite processed by HPT decrease significantly compared to the initial brookite powder. It can be concluded that the electron-hole recombination in the HPT-processed mixture is less than in the initial brookite. SPD processes such as HPT generate various defects in the sample structure, including vacancies, dislocations and grain boundaries [46,50], and these defects have been proven to suppress the recombination of electrons and holes after HPT processing in brookite [44]. Earlier studies using X-ray photoelectron spectroscopy (XPS) confirmed that while the overall valences of titanium and oxygen do not change through HPT processing of $TiO_2$, some oxygen vacancies are formed [44,51]. In addition to XPS, which shows the formation of vacancies on the surface, oxygen vacancies in the subsurface and bulk of HPT-processed brookite were also confirmed earlier by Raman spectroscopy and electron spin resonance, respectively [44]. It should be noted that a high HPT processing temperature can enhance the formation of vacancies; however, high temperatures were avoided in this study due to the presence of PP plastics. Instead, a larger external strain was applied in this study by increasing the number of turns to $N = 3$, which is known to be effective in enhancing the fraction of defects [38,39]. To further elucidate whether mixing with PP has any extra effect on electron-hole recombination, both pure brookite and brookite-PP mixtures were processed by HPT for 1 turn, and their photoluminescence spectra were compared. As shown in Fig. 3d, under similar HPT conditions, the PL intensity decreases from 200 cps to 50 cps with the addition of PP. This reduction indicates a significant decrease in electron-hole recombination in the mixture of brookite and PP plastic due to the transfer of charge carriers between brookite and PP. Similar to organic-inorganic composite photocatalysts, PP (organic molecule) is likely excited, and its excited electrons are injected into the conduction band of brookite (inorganic semiconductor), leading to longer charge separation and slower recombination [52-55]. These PL measurements suggest that brookite-PP mixed by HPT has a higher potential for photoreforming.

A comparison of the hydrogen production using the brookite-PP mixture with and without HPT, along with the blank test, is presented in Fig. 4. The blank tests, either without catalyst under irradiation (curve from 0 to 20 h) or with catalyst in the absence of light (datum at time zero), show no hydrogen production without photoreforming. For the sample without HPT processing ($N = 0$), the amount of hydrogen production increases with the increase of irradiation time until 4 h and subsequently stops at a level of 0.5 mmol $g^{-1}$. The amount of hydrogen production increases to 2.0 mmol $g^{-1}$ and 2.7 mmol $g^{-1}$ after HPT processing for $N = 1$ and 3 turns, respectively (4-5 times enhancement). It should be noted that the standard deviation of gas amount measurements by gas chromatography for three different tests was less than 10%, confirming the reliability of the gas chromatograph system. Moreover, three independent tests for PP mixed with brookite without HPT processing, shown in Fig. 4, indicate an average error of less than 10%, confirming the reproducibility of the photoreforming data. The increase in hydrogen production from $N = 1$ to $N = 3$ is due to the enhancement of the externally applied shear strain because the shear strain is proportional to the number of HPT turns ($\gamma = 2\pi rN/h$; $\gamma$: external applied shear strain; $r$: distance from the rotation center; $N$: number of rotations; $h$: height of the sample) [56,57]. It is well established that the fraction of HPT-induced defects increases and phase mixing improves with increasing shear strain [56,57], which should be responsible for the higher activity of the sample processed with $N = 3$. A larger number of turns beyond $N = 3$ is expected to be even more effective, although a saturation of lattice defects usually occurs at high shear strains where no change in the microstructure occurs with further increases in the number of turns [57]. Increasing the number of turns can be technically challenging due to possible damage to the HPT anvils by hard brookite ceramics; thus, rotations were limited to $N = 3$ in this study. It is noted that after a certain reaction time (4 h, 22 h, and 24 h for $N = 0$, $N = 1$, and $N = 3$, respectively), the production of hydrogen stops due to the consumption of PP plastics in contact with the catalyst. Since a larger number of turns mixes a more significant amount of PP and catalyst, the hydrogen production at the plateau region is the most enhanced for $N = 3$. The enhancement in photoreforming activity without phase transformation or chemical treatment is promising for $TiO_2$ photocatalysis [33-37].

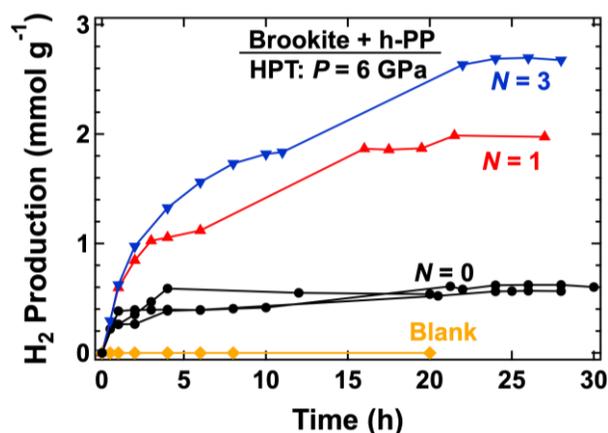

Figure 4. Hydrogen production from photoreforming of PP plastic versus irradiation time for brookite-PP mixtures without HPT ($N = 0$) and with HPT processing for $N = 1$ and 3 turns, including blank test. Three independent tests were conducted for $N = 0$ to examine the reliability of the photoreforming system and the reproducibility of data.

Since the photoreforming process not only generates hydrogen from water in the reduction half-reaction but also produces various organic materials from PP in the oxidation half-reaction (Fig. 1a), GC-MS was used to identify the organic products. It should be noted that the identification of products was conducted using standard peaks provided by the NIST (National Institute of Standards and Technology) mass spectral search program available in the authors' GC-MS machine. The possible chemical compounds and the degradation pathway detected by GC-MS are summarized in Table 1. The oxidized products in Table 1 are considered to have low toxicity to human health and the aquatic environment according to the Hazardous Substances Data Bank of the International Agency for Research on Cancer and the National Institute of Health [58]. Moreover, since they are small molecules, they can be used as initial materials in various chemical reactions. Taken together, simultaneous photocatalytic hydrogen production and PP plastic degradation is achievable using brookite as a catalyst and HPT as a pre-catalysis treatment. Such a process applies not only to PP but also potentially to a wide range of thermoplastics that have applications in various economic sectors [1-12]. Despite these promising results, future studies are needed to verify the long-term applicability of the current catalytic system because cycling performance is considered a main requirement for commercializing photocatalysts such as $TiO_2$ [33,34].

## 4. Discussion

Here, two critical points warrant further discussion: (i) the underlying mechanism behind the simultaneous photocatalytic hydrogen production and PP plastic degradation and (ii) the reasons for the superior photocatalytic performance of the mixture of brookite and PP after HPT processing.

Regarding the first issue, during the photoreforming process under light irradiation, electrons in the photocatalyst are excited to the conduction band ($e^-_{CB}$), where they act as reducing agents to reduce water into hydrogen. Meanwhile, PP plastic serves as a sacrificial electron donor. The photogenerated holes in the valence band ($h^+_{VB}$) function as oxidants, promoting the oxidation of PP plastics into various organic molecules. The reactions involved can be summarized as follows [21-23]:

Table 1. Organic compounds produced from PP plastic and possible degradation pathway for brookite-PP mixture processed by HPT for $N = 3$, subjected to photoreforming and analyzed by GC-MS. Arrows show the degradation pathway from plastic waste to small molecules.

| Retention Time (min) | Compound Name | m/z for Main Ions | Formula | Possible Structure |
|---|---|---|---|---|
| | h-PP | | | 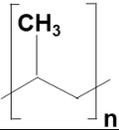 |
| | |  | | |
| 2.465 | 3-Oxapentanol-1, 4-[(2,3-dimethyl)phynyl] | 59, 75, **105** | $C_{12}H_{18}O_2$ | 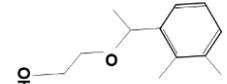 |
| 2.465 | 2-(2-Methoxyethoxy)ethyl Benzoate | 59, 75, **105** | $C_{12}H_{16}O_4$ | 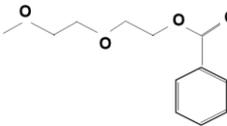 |
| 2.585 | Methoxyacetic acid, benzyl ester | 45, 61, **191** | $C_{10}H_{12}O_3$ | 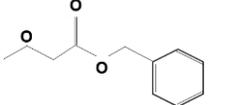 |
| 2.585 | 2-O-Benzyl-d-arabinose | 45, 61, **191** | $C_{12}H_{16}O_5$ | 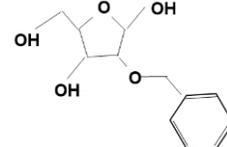 |
| 2.585 | Ethanol, 2-[2-phenylmethoxy]ethoxy] | 45, 61, **191** | $C_{11}H_{16}O_3$ | 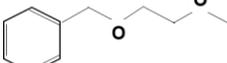 |
| 2.585 | Spiro[2,4]hepta-4,6-diene | 45, 61, **191** | $C_7H_8$ | 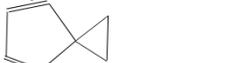 |
| 2.585 | 2-Benzyloxy-3-methyl-1,4-butanediol | 45, 61, **191** | $C_{12}H_{18}O_3$ | 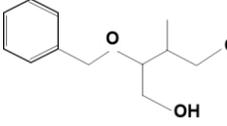 |
| | | 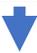 | | |
| 2.115 | Acetic acid, butoxyhydroxy-, butyl ester | **41**, 73, 86 | $C_{10}H_{20}O_4$ | 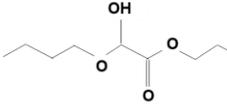 |
| 8.865 | Tridecyl acrylate | 73, 97, **135** | $C_{16}H_{30}O_2$ | 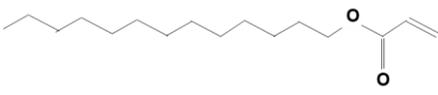 |
| 8.865 | 4-Dodecanol | 73, 97, **135** | $C_{12}H_{26}O$ | 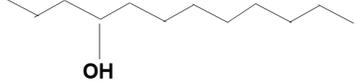 |
| 8.865 | 2-Methyl-1-Pentadecene | 73, 97, **135** | $C_{16}H_{32}$ | 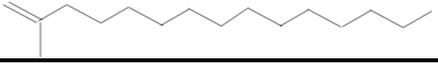 |

$$\text{Brookite} + h\nu \rightarrow h^+_{VB} + e^-_{CB} \qquad (1)$$
$$2H^+ + 2e^- \rightarrow H_2 \qquad (2)$$
$$h^+_{VB} + OH^- \rightarrow \bullet OH \qquad (3)$$
$$h^+_{VB} / \bullet OH + PP \rightarrow \text{Organic Products} \qquad (4)$$

Since all these reactions occur only by renewable solar energy [24-32], the photoreforming process is considered a cleaner technology compared to other plastic degradation methods [13-21]. Although these reactions suggest that both holes and hydroxyl radicals are involved in the transformation of PP, future quenching tests using scavengers are needed to determine which species contribute more significantly to the reactions [59].

Regarding the second issue, several factors contribute to the efficient photoreforming of the mixture of brookite and PP plastic after HPT processing. First, as reported in a previous study, HPT produces defects, such as oxygen vacancies, $Ti^{3+}$ radicals, dislocations, and nanograin boundaries in brookite [44]. These defects serve as active sites for the reaction and facilitate the separation and migration of electron-hole pairs, thereby enhancing photocatalytic activity [60-62]. Second, the HPT process creates defects such as cracks in PP, forms nano-sized brookite, and increases the contact area between brookite, PP plastic, and the aqueous solution. Third, HPT transforms the brookite-PP mixture into a composite that can function as an inorganic-organic composite photocatalyst. Charge carrier transfer at the interphases of such photocatalysts enhances charge separation and diminishes recombination, as evidenced by the PL data. These factors collectively contribute to the superior photocatalytic activity of the HPT-processed brookite and PP plastic mixture, suggesting a new functional application for SPD processing [45,63] which extends the earlier reported photocatalytic applications of severely deformed materials [44,51,64].

## 5. Conclusions

This study introduces a new route for the photoreforming of PP plastics and shows the impact of the HPT method as the pre-catalysis stage of this photoreforming process. Simultaneous water splitting to hydrogen and PP plastic degradation to a variety of organic compounds are achieved. The photoreforming performance systematically improved with an increasing number of HPT turns, due to the formation of defects in both plastics and catalysts and the formation of abundant inorganic-organic interphases. These findings suggest a new potential application of SPD processing for plastic waste management.

**Acknowledgments**

This study is supported partly by Mitsui Chemicals, Inc., Japan, and partly through a Grant-in-Aid from the Japan Society for the Promotion of Science (JP22K18737).